\documentclass{article}
\usepackage[utf8]{inputenc}
\usepackage[T1]{fontenc}
\usepackage{authblk}   
\usepackage{multirow}
\usepackage[a4paper,top=1.5cm,bottom=2cm,left=2cm,right=2cm]{geometry} 
\usepackage{hyperref}
\usepackage{amsmath}
\usepackage{graphicx}
\usepackage{subcaption}
\usepackage{float}

\title{\vspace{-2cm}On the RAID dataset of perceptual responses: analysis and statistical causes}

\author[1]{Paula Daudén-Oliver\thanks{corresponding author: paula.dauden@uv.es}}
\author[2,3]{David Agost-Beltran}
\author[2,3]{Emilio Sansano-Sansano}
\author[2,3]{Raul Montoliu}
\author[1]{Valero Laparra}
\author[1]{Jesús Malo}
\author[2,4,5]{Marina Martínez-Garcia}

\affil[1]{Universitat de València, Image Processing Laboratory, Valencia, 46980, Spain.}
\affil[2]{Universitat Jaume I, Castellón, 12071, Spain.}
\affil[3]{Institute of New Imaging Technologies, Universitat Jaume I, Castellón, 12071, Spain.}
\affil[4]{Institut de Matemàtiques de Castelló, Universitat Jaume I, Castelló 12071, Spain.}
\affil[5]{Institut d'estudis feministes Purificación Ecribano, Universitat Jaume I, Castelló 12071, Spain.}

\date{} 
\begin{document}
\maketitle
\begin{abstract}
This work analyzes the RAID dataset to evaluate human responses to affine image distortions, including rotation, translation, scaling, and Gaussian noise. Using Mean Squared Error (MSE), the study establishes human detection thresholds for these distortions, enabling comparison across types. Statistical analysis with ANOVA and Tukey–Kramer tests reveals that observers are significantly more sensitive to Gaussian noise, which consistently produced the lowest detection thresholds. Fourier analysis further shows that high-frequency components act as a visual mask for Gaussian noise, demonstrating a strong correlation between high-frequency energy and detection thresholds. Additionally, spectral orientation influences the perception of rotation. Finally, the study employs the PixelCNN model to show that image probability significantly correlates with detection thresholds for most distortions, suggesting that statistical likelihood affects human visual tolerance.

\end{abstract}

\section{Introduction}

In this work, we analyze human responses to affine distortions and Gaussian noise. For this, we evaluate the responses from RAID-dataset \cite{RAID}, that contains 24 reference images and 864 distorted images (9 levels of distortion). Human responses are measured using the Maximum Likelihood Difference Scaling (MLDS) \cite{MLDS}, obtaining a perceptual scale for each reference image and distortion.

\vspace{18cm}

\section{Perceptual Threshold Estimation}
\begin{figure}[h!]
    \centering
    \includegraphics[width=1\linewidth]{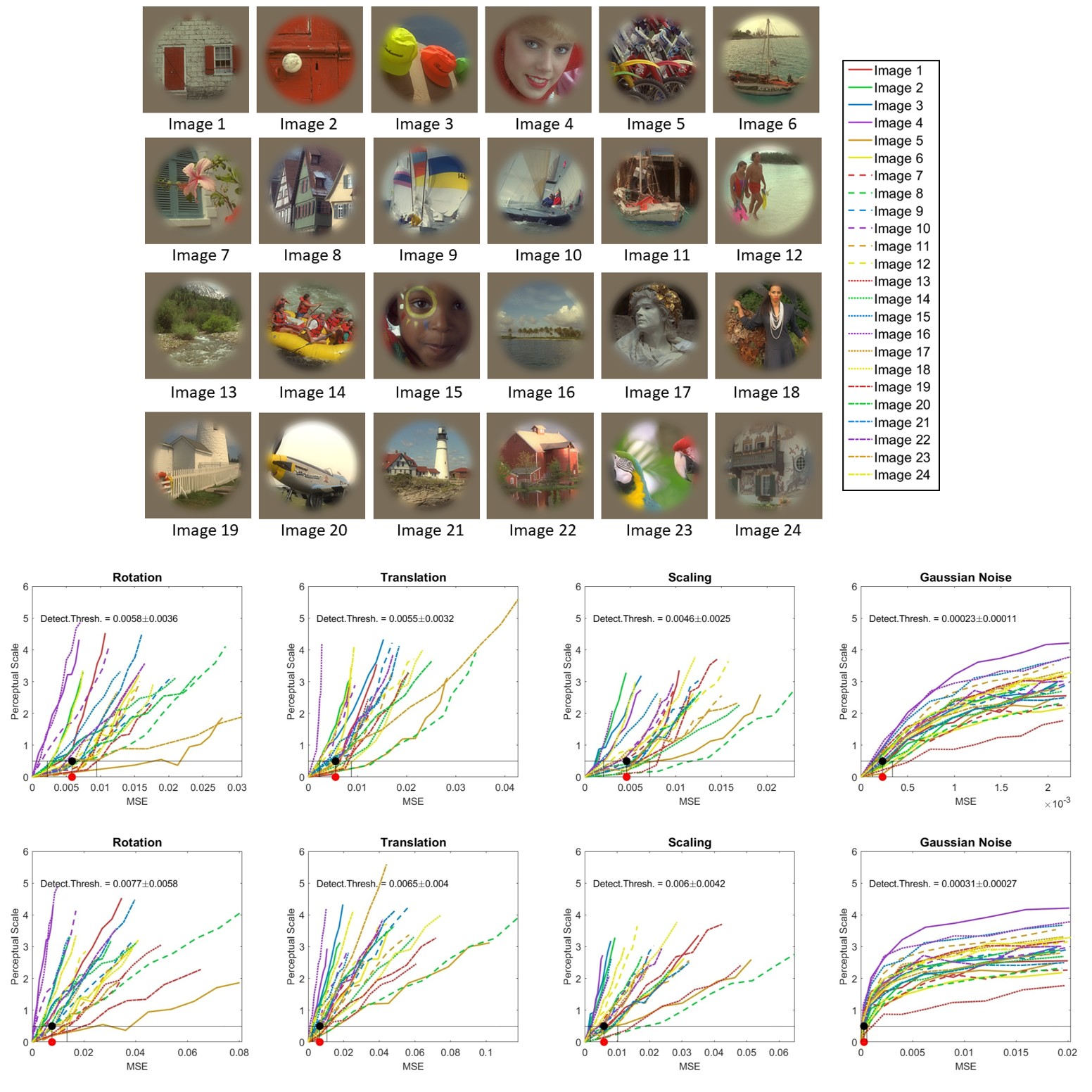}
    \caption{Human thresholds in MSE, calculated between the original and the distorted image (top) and calculated summing the MSE values obtained between consecutive distortion levels, starting from the reference image and the first distorted image (bottom).}
    \label{fig:mse}
\end{figure}

To compare the different types of distortion, the mean squared error (MSE) is computed using two approaches. In the first approach, the MSE is calculated between each distorted image and the corresponding reference image (Figure \ref{fig:mse} top). In the second approach, a cumulative MSE is computed by summing the MSE values obtained between consecutive distortion levels, starting from the reference image and the first distorted image (Figure \ref{fig:mse} bottom). This metric captures the progressive effect of incremental distortions. 
Although no method of measuring MSE perfectly reflects the measurements obtained in the MLDS, because in the experiment comparisons were made between different distortion levels, not only with the original but also with the previous distortion level, the cumulative MSE provides a better reproduction of the experimental measurement.



\begin{figure}[h!]
    \centering
    \includegraphics[width=1\linewidth]{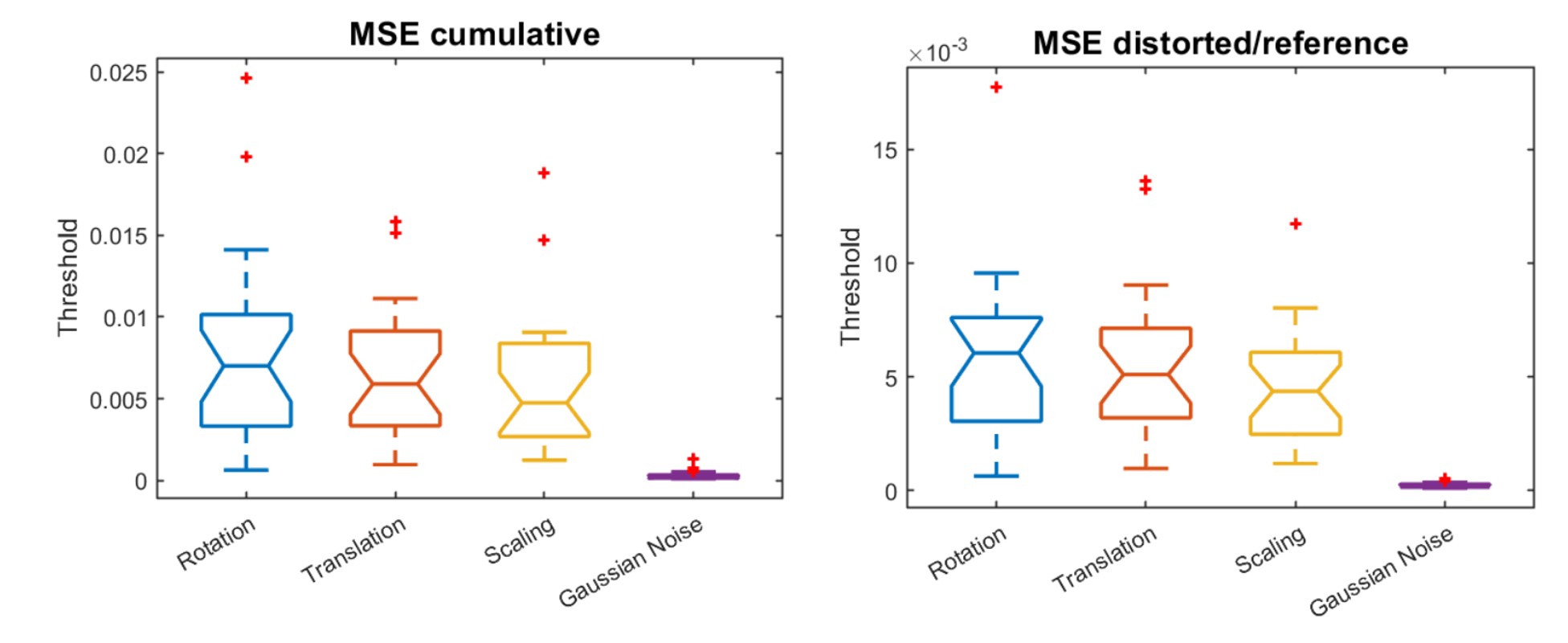}
    \caption{Box plots of the thresholds calculated in MSE, calculated between the original and the distorted image (left) and calculated summing the MSE values obtained between consecutive distortion levels, starting from the reference image and the first distorted image (right).}
    \label{fig:boxplot}
\end{figure}

Figure \ref{fig:boxplot} shows box plots of the thresholds of all images in terms of MSE values: calculated with the reference image (left) and cumulative MSE (right) for all distortions. Gaussian noise produces the lowest thresholds with minimal variability, indicating that we are more sensitive to this distortion. In contrast, rotation, translation, and scaling lead to higher thresholds and a larger deviation, with rotation showing the greatest variability.

Then an ANOVA test was applied to determine whether the thresholds of the different transformations were significantly different, showing that differences exist between groups (p-values $< 0.001$). To identify which groups differ, a Tukey-Kramer test was conducted to pairwise compare the thresholds. In both cases, only the Gaussian noise transformation exhibited a lower threshold than the affine transformations (all p-values $< 0.01$).

A closer inspection of the outliers provides valuable insight into the influence of image content on human perceptual responses and detection thresholds. For rotation, the outliers correspond to Images 5 and 7, which lack a clear dominant orientation, making rotational distortions less perceptually salient. In the case of translation, Images 5 and 8 exhibit higher thresholds, likely due to their visual complexity or structural repetition, which reduces sensitivity to spatial shifts. For scaling, Images 8 and 13 stand out as outliers, as the absence of well-defined size references in architectural and natural scenes increases tolerance to changes in scale. Finally, for Gaussian noise, Images 5 and 13 present unusually high thresholds, as their high texture content and intrinsic visual variability effectively mask the presence of additive noise.
\vspace{1cm}

To quantify the dependency of detection thresholds on image content, we calculated the Coefficient of Variation ($CV$) for each distortion type. The $CV$ is defined as the ratio of the standard deviation $\sigma$ to the mean $\mu$:$$CV = \frac{\sigma}{\mu}$$This metric provides a normalized measure of dispersion, allowing for a robust comparison across different scales of Mean Squared Error (MSE). Table \ref{tab:cv_results} summarizes the $CV$ for each condition along with their $95\%$ confidence intervals ($CI$). It can be observed that for all distortion types, the variability increases when considering the cumulative MSE. This phenomenon occurs because affine transformations follow a coherent trajectory within the image manifold. Each incremental step moves the image further along a predictable directional path. In contrast, the most significant increase in variability was found in the Gaussian noise (GN) condition. While the thresholds calculated with the reference and distorted image showed the lowest variability ($CV = 0.4569$), the accumulated MSE reached the highest dispersion of all distortions ($CV = 0.8662$). Unlike the directional nature of affine transformations, Gaussian noise represents a high-dimensional random walk in the pixel space. Since each noise injection is independent and stochastic, the total error is the sum of random vectors pointing in multiple directions.

\begin{table}[h]
\centering
\begin{tabular}{lcccc}
\hline
\textbf{Metric Type} 
& \textbf{Rotation} 
& \textbf{Translation} 
& \textbf{Scaling} 
& \textbf{Gaussian noise} \\
\hline
Distorted / Ref. MSE 
& 0.6175 [0.37, 0.81] 
& 0.5806 [0.39, 0.71] 
& 0.5515 [0.38, 0.69] 
& 0.4569 [0.31, 0.56] \\
Cumulative MSE 
& 0.7579 [0.50, 0.94] 
& 0.6191 [0.45, 0.75] 
& 0.7088 [0.45, 0.88] 
& 0.8662 [0.51, 1.06] \\
\hline
\end{tabular}
\caption{Coefficients of Variation ($CV$) and $95\%$ Confidence Intervals for the experimental conditions.}
\label{tab:cv_results}
\end{table}

\begin{figure}[h!]
    \centering
    \includegraphics[width=1\linewidth]{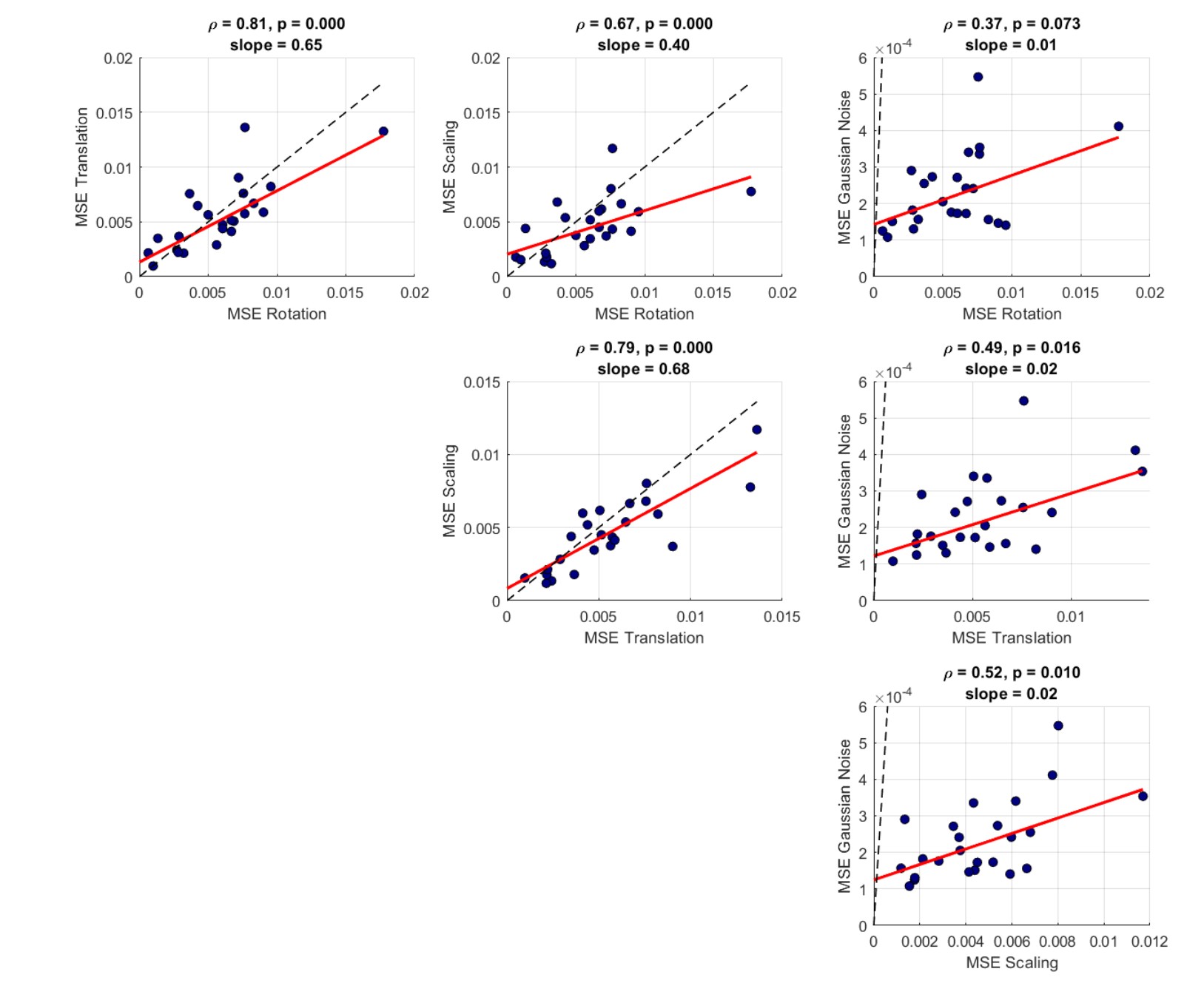}
    \caption{Correlation of human thresholds between different distortions. The red line shows the fitted slope, and the black line represents unity.}
    \label{fig:cor_mse}
\end{figure}

To determine whether the same reference images exhibit higher thresholds across different distortions, Spearman correlation coefficients and slopes between distortions were computed. Figure \ref{fig:cor_mse} shows that affine distortions exhibit stronger relationships, whereas no clear correlation is observed between Gaussian noise and affine distortions. Although ANOVA revealed no significant differences in thresholds among affine distortions, systematic slope deviations from unity were observed between them, indicating that, for the same image, increases in perceptual tolerance to rotation are associated with smaller increases in tolerance to translation and scaling. This pattern suggests a latent perceptual ordering among affine transformations: rotation showing the highest tolerance, followed by translation and scaling, that is not captured by global distribution-level comparisons.

\section{Fourier Analysis}

To analyze the relationship between image content and human responses to the different distortions, each image was characterized in terms of its spectral energy distribution. For this purpose, we computed the two-dimensional discrete Fourier transform of each image and quantified spectral energy directly in the frequency domain.

\begin{figure}[h!]
    \centering
    \includegraphics[width=1\linewidth]{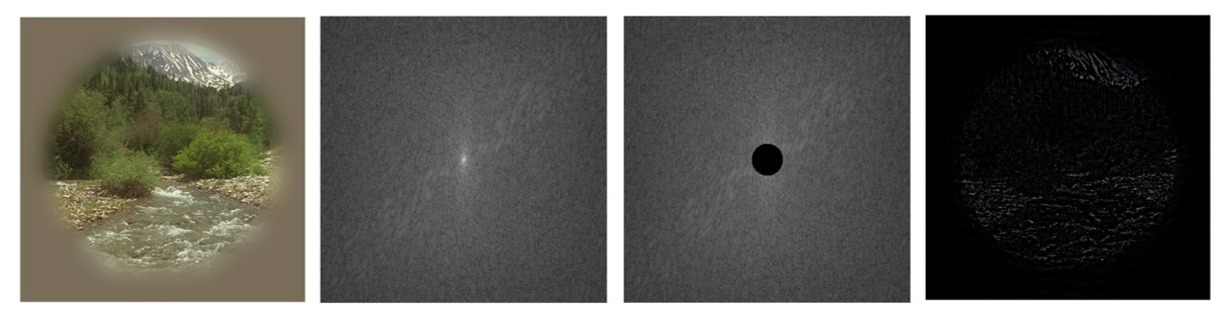}
    \caption{Circular filter applied to the images. From left to right:1. Example of a reference image. 2. Reference image in the Fourier domain. 3. Filter applied to the image in the Fourier domain. 4. Image filtered.}
    \label{fig:ej_circ}
\end{figure}

In a first analysis, a circular high-pass filter was applied in the Fourier domain to isolate the high spatial frequency components of each image (Fig \ref{fig:ej_circ}).

Then, the proportion of high-frequency energy was computed as the ratio between the high-frequency energy and the total image energy. This measure was used to analyze the correlation between the energy proportion and the human responses for each image, as well as with the perceptual thresholds.

\begin{figure}[h!]
    \centering
    \includegraphics[width=1\linewidth]{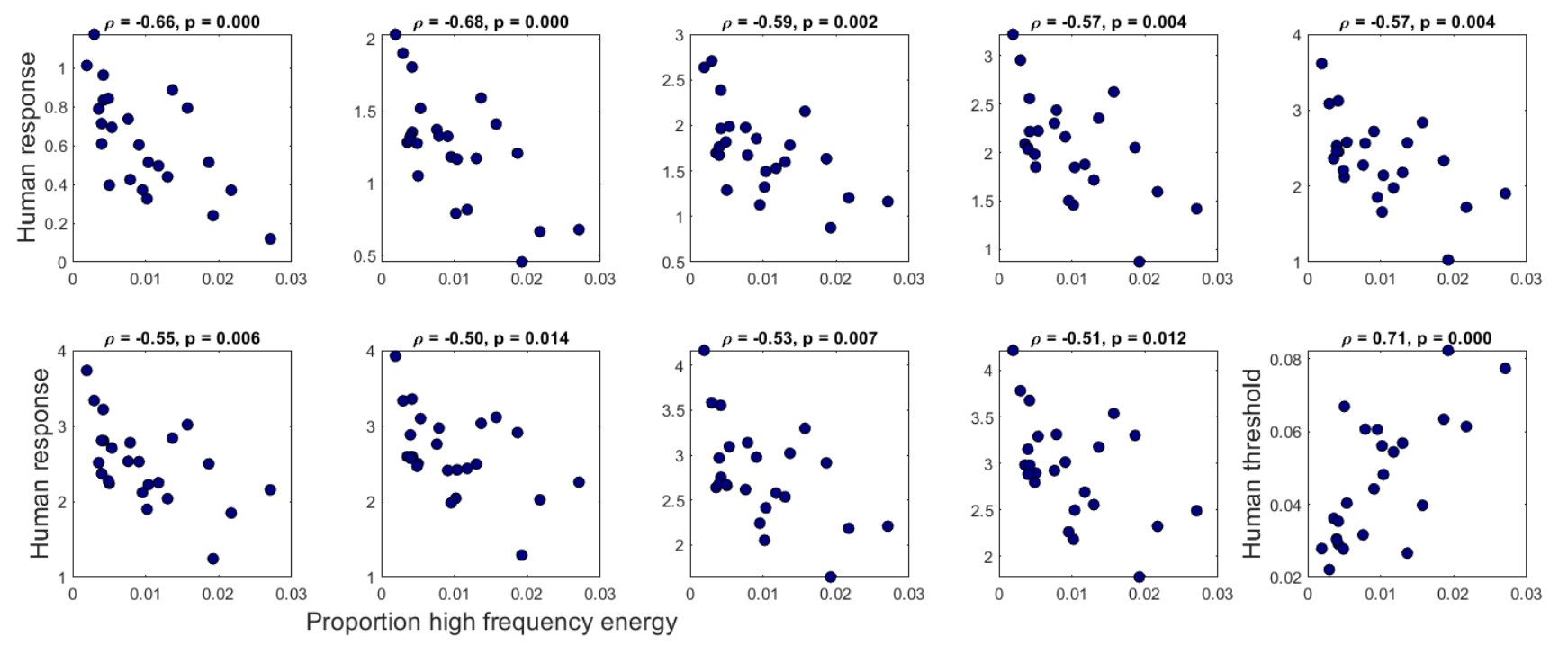}
    \caption{Pearson correlation between the proportion of high frequency energy and the responses to Gaussian noise for each level of distortion and for the threshold derived from the responses.}
    \label{fig:cor_resp}
\end{figure}

Figure \ref{fig:cor_resp} illustrates the correlation between the high-frequency proportion and human responses across the nine levels of distortion. A clear inverse correlation is observed, indicating that Gaussian noise distortions are less perceptible in images with a higher proportion of high-frequency components. Furthermore, the correlation between image thresholds and this proportion was found to be 0.71, demonstrating a strong relationship between these two variables.

\begin{figure}[h!]
    \centering
    \includegraphics[width=1\linewidth]{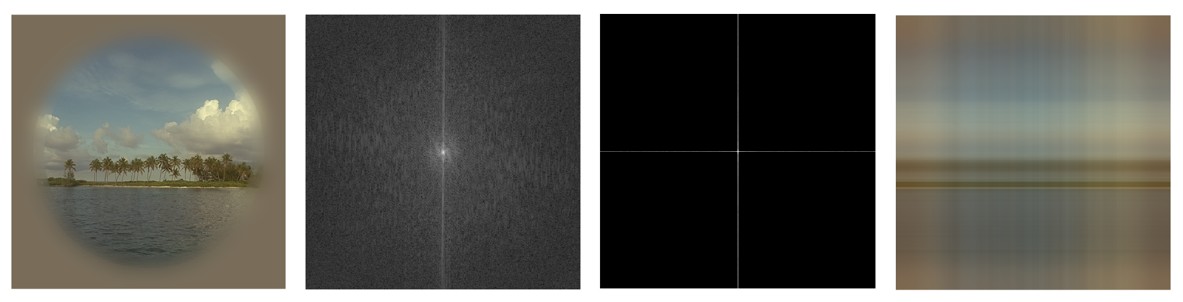}
    \caption{{Horizontal and vertical filter applied to the images. From left to right:1. Example of a reference image. 2. Reference image in the Fourier domain. 3. Filter applied to the image in the Fourier domain. 4. Image filtered.}}
    \label{fig:filt_cruz}
\end{figure}

\begin{figure}[h!]
    \centering
    \includegraphics[width=1\linewidth]{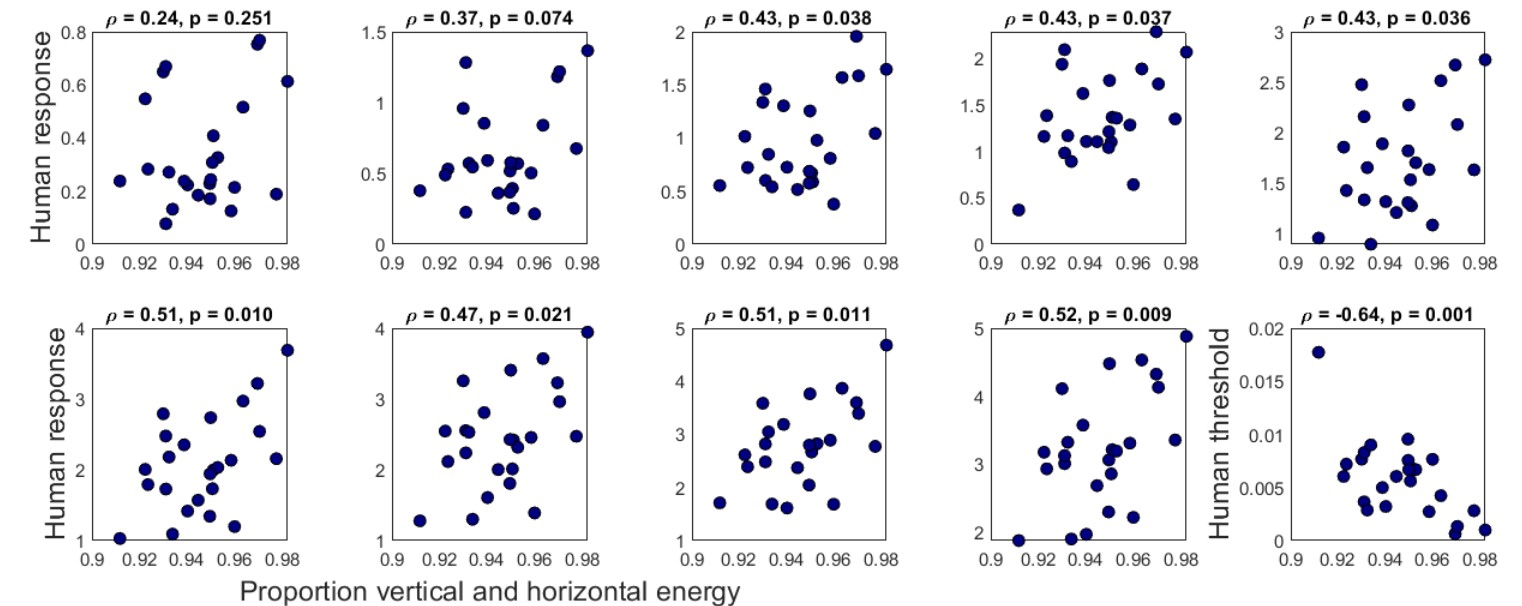}
    \caption{Pearson correlation between the proportion of vertical and horizontal energy and the responses to Rotation for each level of distortion and for the threshold derived from the responses.}
    \label{fig:cor_rot}
\end{figure}

A second filter was also applied, which retains only the vertical and horizontal components of the image (Figure \ref{fig:filt_cruz}). As before, the proportion of vertical and horizontal energy was computed for each image. Figure \ref{fig:cor_rot} presents the correlation between human responses to rotation at each distortion level and the proportion of horizontal and vertical energy. At lower distortion levels, no clear correlation is observed, which may be related to the reduced perceptual sensitivity to small affine transformations, as discussed previously. At higher distortion levels, a moderate correlation (r = 0.52) is observed, suggesting a potential relationship between the proportion of horizontal and vertical energy and human perceptual responses to rotation. Additionally, a negative correlation (r = -0.64) is observed for the detection thresholds, indicating greater difficulty in detecting rotational distortions when images lack strong horizontal or vertical components. When the same analysis is conducted for the remaining distortions, no meaningful correlation is found, which is consistent with the fact that these transformations are not directly related to image orientation.


\begin{figure}[h!]
    \centering
    \begin{subfigure}{1\linewidth}
        \centering
        \includegraphics[width=0.9\linewidth]{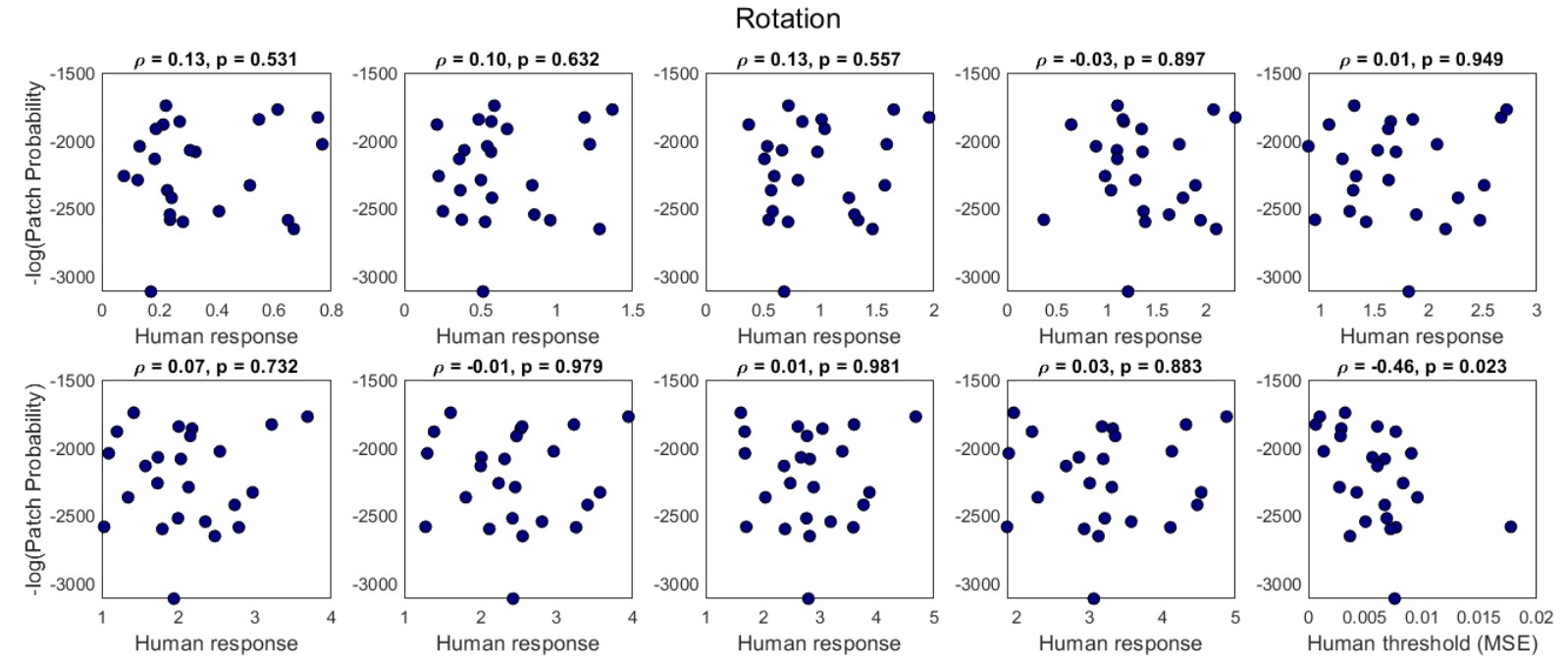}
    \end{subfigure}
    \begin{subfigure}{1\linewidth}
        \centering
        \includegraphics[width=0.9\linewidth]{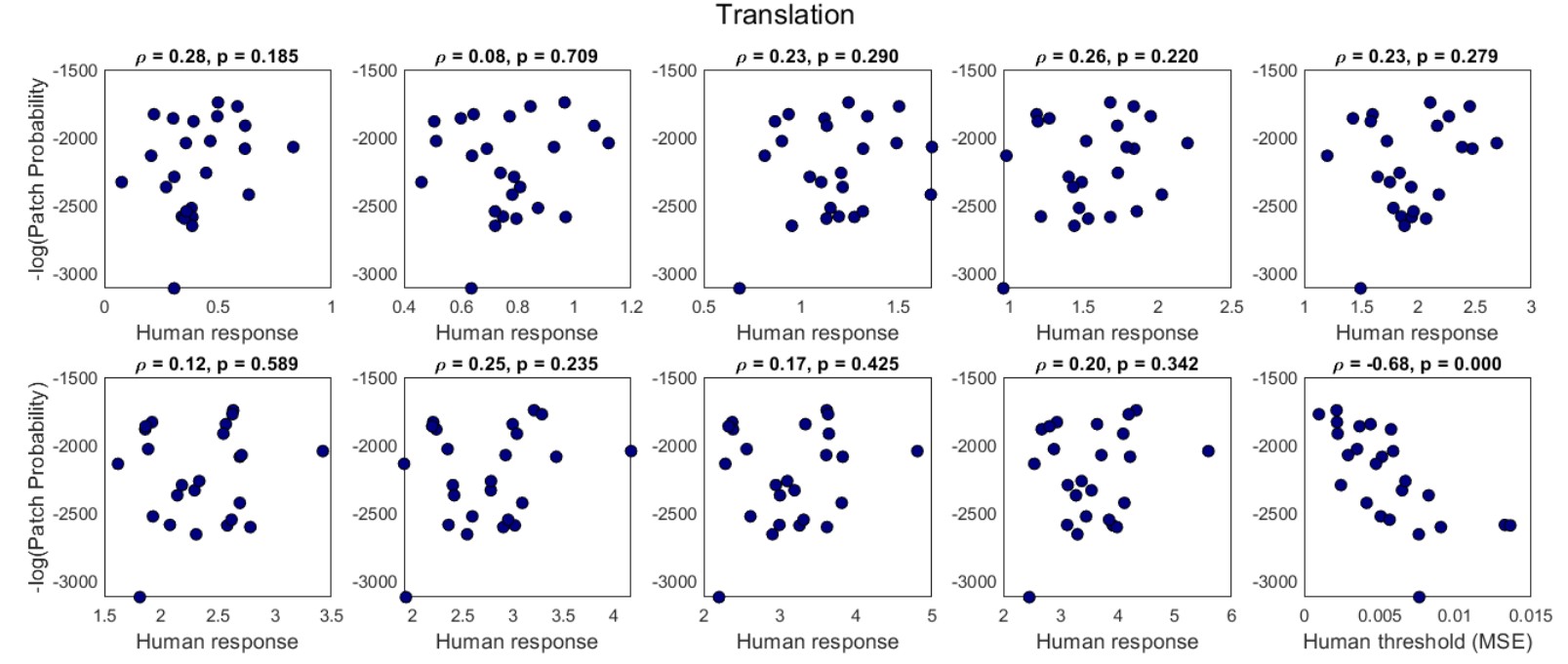}
    \end{subfigure}
    \begin{subfigure}{1\linewidth}
        \centering
        \includegraphics[width=0.9\linewidth]{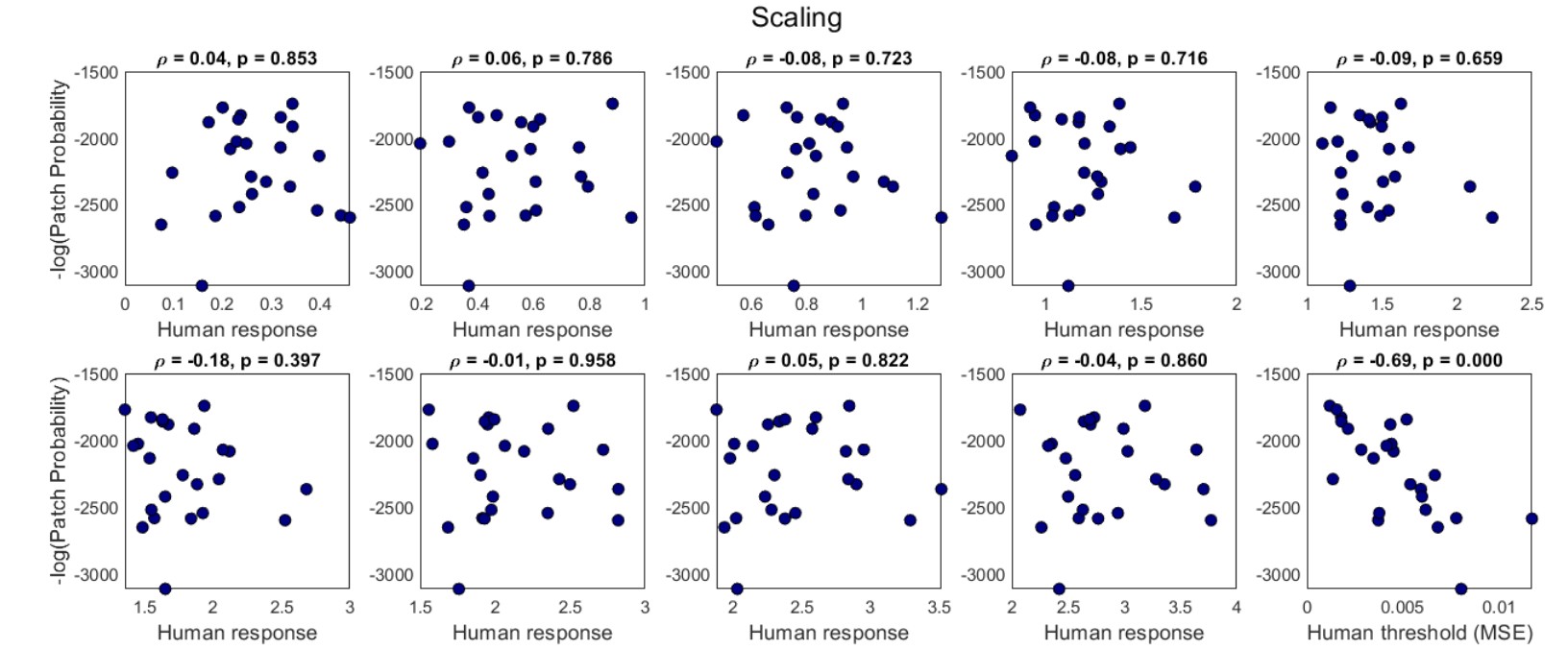}
    \end{subfigure}
    \begin{subfigure}{1\linewidth}
        \centering
        \includegraphics[width=0.9\linewidth]{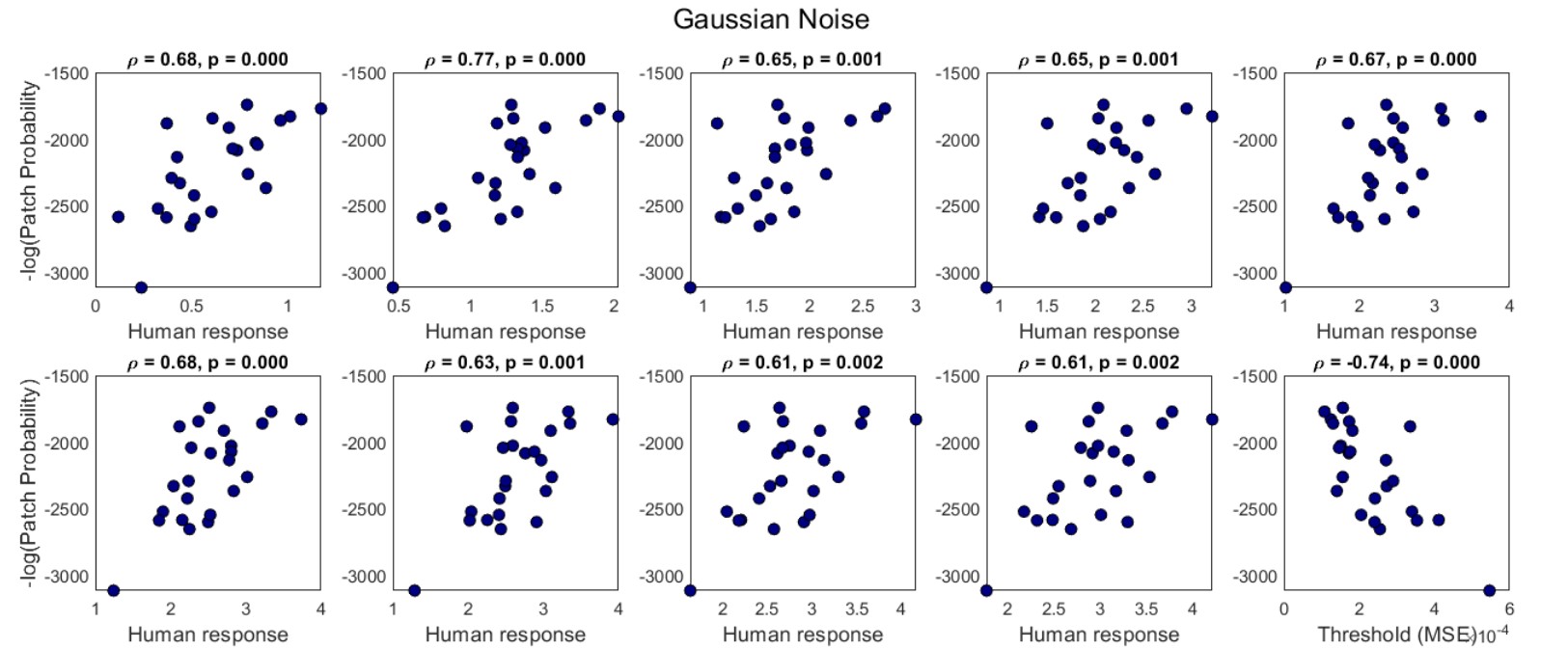}
    \end{subfigure}
    \caption{Correlation between the probabilities of the reference images and the human responses in each level distortion and thresholds.}
    \label{fig:prob_ref}
\end{figure}

\vspace{5 cm}
\section{Probabilities}
To evaluate whether humans exhibit higher sensitivity to distortions in more probable images, reference images were analyzed using the PixelCNN model \cite{PixeCNN}. The model processes 32 × 32 image patches and computes the probability of each patch. A single probability value per image was then obtained by averaging the probabilities across all patches.

Figure \ref{fig:prob_ref} illustrates the correlation between image probability and both human responses and distortion detection thresholds. The results reveal a strong correlation between image probability and detection thresholds for translation, scaling, and Gaussian noise, as well as a moderate correlation for rotation. In contrast, human responses show a significant correlation with image probability only in the case of Gaussian noise.

\section*{Acknowledgements}

The work was partially funded by the Spanish Government and the EU under the  MCIN/AEI/FEDER/UE Grant PID2023-152133NB-I00,  and by the BBVA Foundations of Science program in Maths, Stats, Comp. Sci. and AI, grant VIS4NN: \emph{Vision Science for Artificial Neural Networks}.

\bibliographystyle{plain}
\bibliography{bib_raid}

\end{document}